# The Impact of Cost and Network Topology on Urban Mobility: A Study of Public Bicycle Usage in 2 U.S. Cities

Raja Jurdak[1,2]*

1 Commonwealth Scientific Industrial and Research Organisation, Brisbane, QLD, Australia, 2 University of Queensland, St. Lucia, Brisbane, QLD, Australia

## Abstract

Understanding the drivers of urban mobility is vital for epidemiology, urban planning, and communication networks. Human movements have so far been studied by observing people's positions in a given space and time, though most recent models only implicitly account for expected costs and returns for movements. This paper explores the explicit impact of cost and network topology on mobility dynamics, using data from 2 city-wide public bicycle share systems in the USA. User mobility is characterized through the distribution of trip durations, while network topology is characterized through the pairwise distances between stations and the popularity of stations and routes. Despite significant differences in station density and physical layout between the 2 cities, trip durations follow remarkably similar distributions that exhibit cost sensitive trends around pricing point boundaries, particularly with long-term users of the system. Based on the results, recommendations for dynamic pricing and incentive schemes are provided to positively influence mobility patterns and guide improved planning and management of public bicycle systems to increase uptake.

**Funding:** The work of Jurdak is funded by the Commonwealth Scientific Industrial and Research Organisation (CSIRO), Australia. The funders had no role in study design, data collection and analysis, decision to publish, or preparation of the manuscript.

**Competing Interests:** The author has declared that no competing interests exist.

* E-mail: rjurdak@ieee.org

## Introduction

Increasing greenhouse gas emissions and traffic congestion have driven large cities to build public shared bicycle systems as an active, low-emissions, and sustainable form of transport. City councils typically install bicycle stations at hundreds of locations across the city, and citizens can then use any available bicycle from these stations. Public bicycles are seen as the last mile of transportation systems [24,26], with the potential to bridge the gap between people's home and workplace and other transportation options, and they are experiencing exponential growth in large cities in Europe, Asia, North America and Australia [1,24,26].

While these systems have attracted large public and private investments and media hype, their impact on gas emissions has been limited due to their failure to attract car commuters in significant numbers to switch to public bicycles [1,25,29]. Fundamental to improving uptake of public bicycle share systems and to rendering them financially sustainable in the long-term is to understand the factors that affect their usage patterns.

Recent work on understanding human mobility has characterized the distance people travel and the duration of their trips as power law distributions, i.e. people tend to mostly travel for short periods and distances, with longer trips being exponentially less likely [2,3,4,16]. While highly valuable, power law characterization does not capture the associated costs and returns of mobility. The drivers of all forms of human mobility, spanning short-distance journey to work and residential mobility to long distance migration within and between countries [5], are established to be a function of cost and return [6,7], both monetary [8] and non-monetary [9]. Gravity models [10,11], where more populous locations are more attractive for incoming trips, implicitly assume that mobility costs are proportional to distance. More recent work has linked mobility decisions to other incentives, such as employment options [12].

Kolbl and Helbing [27] have pointed out the existence of a daily energy budget for urban movement, but they did not explicitly quantify how monetary cost interacts with this energy budget. In general, the impact of monetary cost has not been quantified for most types of mobility, partly because of the reliance on available data sources, such as census records, mobile phone traces, and GPS logs that do not directly capture costs and returns associated with mobility. Public transport costs are an exception, as they map specific fares to given routes. The impact of cost and return on mobility using conventional public transport provides only a limited view, as mobility is constrained to fixed routes and thus fails to capture the effects of personal choices. As a result, there is limited understanding of the impact of pricing and incentive structures on mobility dynamics, and the relative importance of these structures compared to other factors, such as the physical layout that governs mobility.

Public bicycle systems provide a unique opportunity to explore the impact of monetary cost on mobility decisions by coupling both the flexibility to choose any route between existing stations and an explicit costing structure. The works in [24,26] provide comprehensive reviews of the history and outlook for public bikesharing, as well as an overview of their evolution and uptake drivers across different regions in recent years. Kolbl and Helbing's work on the energy laws for human travel behaviour [27] across different travel modes provides valuable insights into the usage dynamics of cycling. However, it did not consider the current generation of public bicycles and how these laws interact with public bikeshare systems. Buck et al. [25] studied the reported trip purpose of public bicycle users in Washington D.C., finding that a large proportion of annual members use the system for utilitarian purposes such as commuting, while casual users tend to use it mostly for tourism.

This paper builds on the above work by exploring the impact of cost thresholds and network topology on how people use the public bikeshare systems. The focus is to characterize the explicit impact of cost on urban mobility and compare its relative importance with network topology through the case study of shared public bicycle systems. Public bicycles typically require users to subscribe to the system or pay a one-off fee before borrowing a bicycle. Nearly all public bicycle systems incentivize shorter trips, where trips of more than 30 minutes are charged extra fees at progressively increasing rates [1,13]. This common costing structure provides an explicit representation of the cost (effective for trips longer than 30 minutes) and return (moving from source to destination).

## Analysis and Results

This paper uses publically available shared bicycle data for Boston [14] and Washington D.C. [15], which include 1000 and 1800 bikes, 100 and 200 stations, and 552,073 and 1,859,773 trips respectively. The data for both cities spans a timeframe from September 2010 to December 2012, where each trip is represented by a start and end station, start and end time, unique bicycle ID, and for the Boston data, demographic information about the user. To check out a bicycle, users must first subscribe to the system. The subscription costs for trips in each of the cities are shown in Table 1, indicating a slight variation in prices between the 2 cities. Once subscribed, additional trips may incur further costs based on trip time (see Figure 1). Only trips of 60 seconds or longer are considered in the analysis to filter out bicycles that are immediately returned without being used. Trips longer than 24 hours are also disregarded as they represent irregular use of the bike-sharing system.

Figure 1 plots the trip cost as a function of time, which is the same for the public bicycle systems in both Washington D.C. and Boston. Trips below 30 minutes incur zero additional cost (maximum return) beyond the initial cost of becoming a member, while longer trips incur an increasing cost at half hour intervals, with the maximum cost at trips of between 7 and 24 hours capped at $100 for casual users and $80 for registered users. Registered users are defined as having either a monthly or annual membership, with all other users classified as casual.

### Cost Impact

To explore the impact of cost structure on public bicycle usage, Figure 2 plots the distribution of trip durations in these cities. The trip duration distributions for both cities are remarkably similar with a strong bias towards shorter trips, and an apparent power law decrease beyond a certain duration, which confirms observations in earlier work around trip duration [17] and energy [27]. Within the cost-free period, there is a tendency towards shorter trips with a peak around 6 minutes. A fairly broad spread of trip times within 30 minutes is observed, with a sharp decline in the likelihood for trips just under and above 30 minutes. This captures the behavior of most public bicycle users to try to maximize their travel distance and time (mobility return) without increasing their cost. Once they have incurred the cost ($2) by exceeding the 30-minute mark, most users tend to keep the bicycle to increase their return for cost they have already incurred. The distinct decline in trip time around 30 minutes continues until about 50 minutes,

**Table 1.** One-time subscription costs in each of the 2 cities.

| City | Annual ($) | Monthly ($) | Daily ($) | 72-hour ($) |
|---|---|---|---|---|
| Boston | 85 | 20 | 6 | 12 |
| Washington D.C. | 75 | 25 | 7 | 15 |

Individual trip costs depend on the trip time and are shown in Figure 1.
doi:10.1371/journal.pone.0079396.t001

when users realize they may incur a significantly higher cost ($6) if they do not return bicycles within a few minutes. A minor bump appears in the plot between 50 and 60 minutes confirming this behavior. Beyond 60 minutes, where the cost of every additional half hour is the same at $8, trip times exhibit a power-law distribution of the form:

$$y = ax^k$$

in both cities (Boston: $a = 2.65 \times 10^5$; $k = -2.4156$; D.C.: $a = 4.28 \times 10^5$; $k = -2.4643$), where longer trips are exponentially less likely than shorter trips. The deviations from the main power law trend around the cost boundaries of 30 and 60 minutes, with a tendency to cut trips just before 30 minutes and to extend them further when they exceed 30 minutes, suggest that users are adapting their trip times to minimize cost.

To further explore this effect, Figure 3 shows the trip duration distribution separately for casual and registered users in Boston. Casual users, whose primary primary use of the bicycles is for tourism [25], clearly take longer trips on average (3283 seconds) than registered users (818 seconds). Nevertheless, the cost sensitivity for casual users around the 30 minute and 60 minute marks persists with the small bumps and troughs before and after the pay boundaries respectively. Beyond simply taking shorter trips on average, registered users appear to have a much higher cost-sensitivity around the 30 minute mark, with a visible drop in the slope of the distribution shortly before this limit of the cost-free period. This steeper slope remains stable even for longer trips. The higher cost sensitivity of registered users may be a result of their closer knowledge of the bikesharing system, its spatial layout, and the cost structure, which enables them to optimize their use of the system without incurring additional cost.

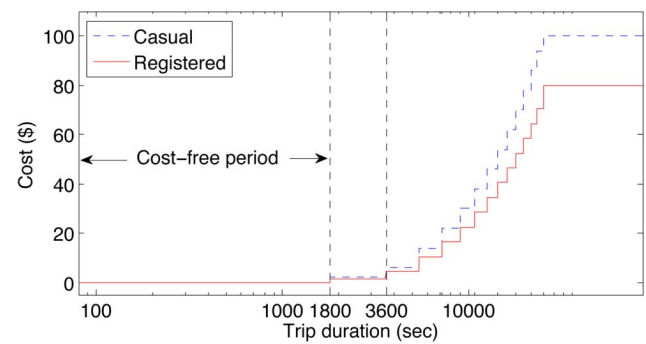

**Figure 1. The monetary cost of public bicycle trips as a function of time.** The cost is the same in both cities.
doi:10.1371/journal.pone.0079396.g001

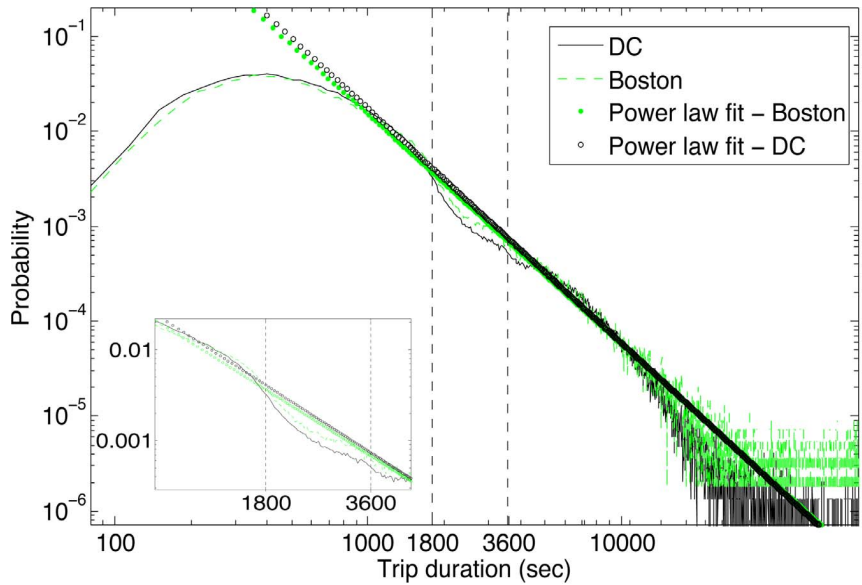

**Figure 2. Distribution of trip durations in Boston and Washington, D.C.** clearly shows a broad spread of trips durations within the cost-free period and a more common power-law distribution for trips longer than 1 hour.
doi:10.1371/journal.pone.0079396.g002

### Topology Impact

Having established a strong similarity in the trip duration distributions and the cost sensitivity of trip durations around the price boundaries, the dependence of trip durations on station topology is now explored. The high similarity in trip duration distributions between the two cities suggests that, if station topology is a strong determinant of trip durations, the topologies in Boston and Washington should also be similar.

Figure 4 shows the distribution of inter-station distances in both cities, as a measure of the spatial station density. Boston stations are on average located much closer to other stations (M:113.09, S:82.21) compared to Washington D.C. (M:247.32,S:162.69). In other words, a bicycle trip in Boston encounters many more stations before reaching its destination relative to a trip in Washington D.C. with comparable distance and duration. Despite the spatial and topological differences, which have been shown to impact trip distributions at finer spatial scales [17], the global trip distributions in the two cities remain highly similar. The nearly identical trip duration distributions coupled with the highly heterogeneous station distributions between the two cities indicates that the (non-monetary) energy cost [27] is the most likely driver of this trip distribution. As Kolbl and Helbing report in their long-term study, there appears to be a universal energy budget for daily travel that is independent of transportation mode. They also highlight that average bicycle trips are around 42 minutes. In comparison, Figure 3 indicates an average trip time of around 14 minutes for registered users. This apparent underuse of the daily energy budget for registered users suggests that they are using the system in conjunction with other forms of transport, which would support the last mile hypothesis [24,26]. Casual users, on the other hand, have average trip times at around 54 minutes, which may arise from the increased frequency of pauses during "touristic" trips undertaken by these users. These pauses are most likely not accounted for in the daily travel energy budget that is dominated by commuters. The consistent deviation in trip durations of public bicycle users around the pricing boundaries in the two cities appears to be primarily due to cost sensitivity rather than to spatial considerations, such as specific station locations.

An alternative explanation to the remarkably similar trip durations in Boston and Washington D.C., including the cost sensitivity around pricing boundaries, is that a few hub stations dominate the bicycle usage, and the durations of the trips among these hub stations may be causing a bias in the trip duration distribution. To explore this possibility, the station popularity for each station is computed as the number of trips that start or end at that station, in order to generate a ranking of all the stations in descending order according to their number of trips. Each station is then assigned a popularity rank, where the most popular station in a city has a popularity rank of 1, and the $n^{th}$ most popular station has a rank of n. The pairwise dominance of stations is further computed by counting the number of other stations for which each station is the most popular source or destination. This metric is to indicate to what extent specific pairs of stations dominate the trip data.

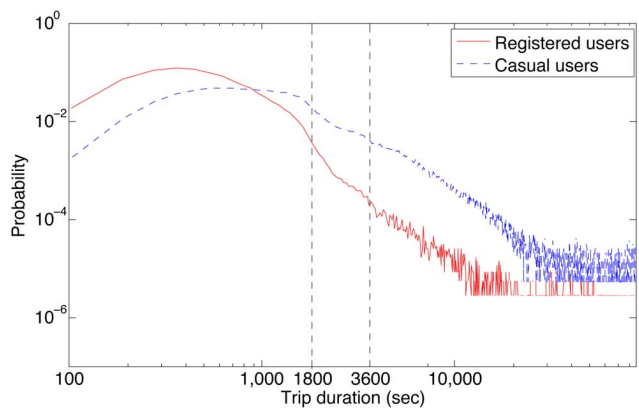

**Figure 3. Distribution of trip durations in Boston split by registered and casual users.** Casual users take longer trips on average, while the trip duration of registered users appears to drop more sharply just before 30 minutes to avoid incurring additional costs.
doi:10.1371/journal.pone.0079396.g003

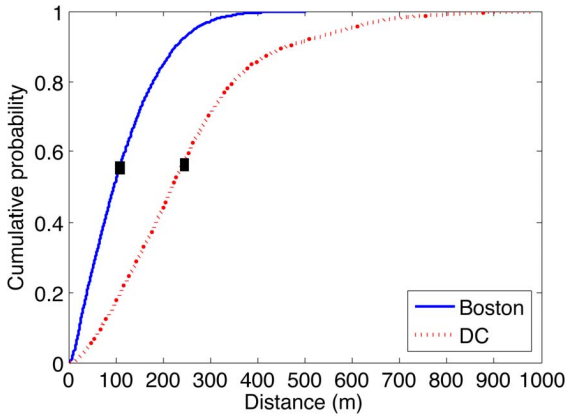

**Figure 4. Cumulative probability distribution of inter-station distances in Boston and Washington D.C. highlights that Boston has a much denser public bicycle network topology.** Black squares indicate mean inter-station distances.
doi:10.1371/journal.pone.0079396.g004

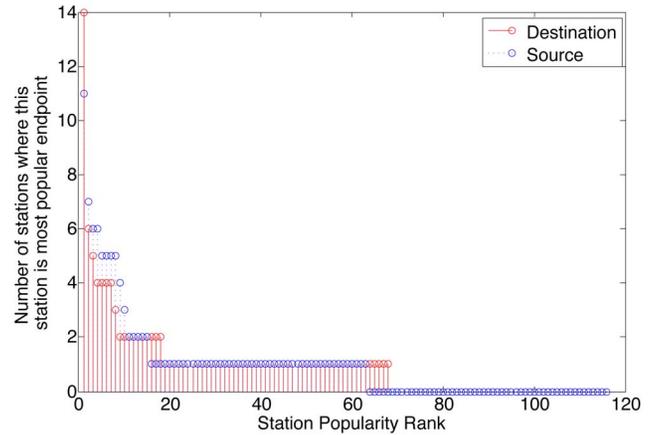

**Figure 6. Station usage patterns show a few popular stations in D.C., and similarly with Boston, the popularity is spread out over more than half the stations.**
doi:10.1371/journal.pone.0079396.g006

Figures 5 and 6 plot the pairwise station popularity versus the station popularity rank for Boston and Washington D.C. respectively. The results confirm previous reports that hub stations exist [18,19], and show that there is asymmetry in popularity as source or destination as a result of commuting patterns. However, these hub stations by no means dominate the usage, with many other stations being used to a slightly lesser extent.

Finally, the dominance of individual routes, represented by source-destination station pairs, and its potential to bias the trip duration distribution is investigated. The number of occurrences for each route are counted and sorted in descending order of occurrences. Every source-destination pair is then assigned a unique route index for each of the two cities, with the results shown in Figure 7. The common feature for both cities is the relative flatness of the plots for the top 1000 routes. In both cities, a handful of routes emerge as the most popular, with 4 most popular routes in Washington D.C. with 2000–3000 trips, and 5 most popular routes in Boston with a similar number of trips. In comparison, all of the top 300 routes in both cities were taken at least 500 times, suggesting that the total trip sample is broadly spread among a large number of routes, and refuting the hypothesis of bias in trip duration distribution due to dominance of a few routes.

## Discussion

The cost sensitivity around pricing boundaries in people's usage patterns supports the current trend for the design of dynamic costing structures or incentive schemes to positively affect the usage of this low emissions means of public transport. Clearly, the time-specific cost-structure of current public bicycle systems has been introduced to contain the issue of excessive borrowing times in previous generations [24]. Dynamic costing and incentives have to therefore strike the right balance between equity in bicycle access, which penalizes longer trips with higher costs, and optimizing demand in target locations, by tactically reducing the cost.

Incentive schemes for using bicycles at the bottom of hills are already in effect in some cities, such as the Velib system in Paris [18,26] through the expansion of the cost-free period to 45 minutes. The incentives recognize that users have to spend extra effort to cycle uphill and gives them an additional 15-minutes of cost-free time to do so. When these additional minutes are not spent in the current trip, they can be saved up for later trips [26].

The idea of "push" and "pull" stations introduces pricing incentives to encourage borrowing or returning bicycles to specific stations [26]. Recent proposals for dynamic public bike sharing systems [24] have gone beyond incentive schemes towards mobile bicycle stations that are relocated based on usage patterns and user demand. Other innovations have included the use of existing street furniture as ad hoc stations and using GPS trackers to allow users to localize the nearest bicycles [26]. A more recent embodiment of these concepts is proposed in the Social Bicycles project [28] that uses regular bike racks with wireless tracking to emulate public bicycles in a distributed and controlled form.

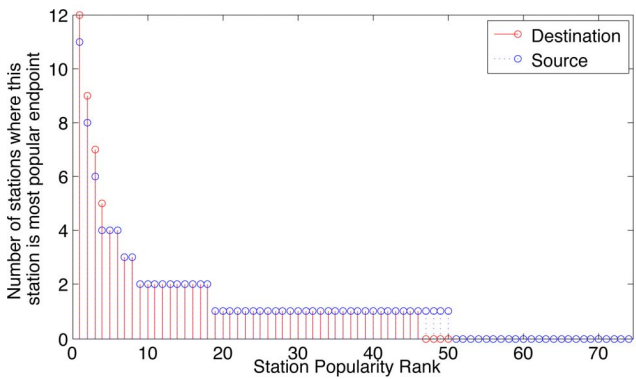

**Figure 5. Station usage patterns show a few popular stations in Boston, yet the popularity of trips is spread out over more than two-thirds of the stations.**
doi:10.1371/journal.pone.0079396.g005

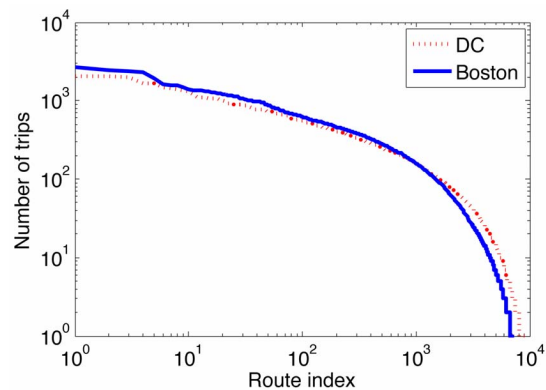

**Figure 7. Number of trips undertaken for unique source-destination pairs reveals a broad spread of trips.**
doi:10.1371/journal.pone.0079396.g007

### General Implications

Given the clear cost sensitivity of trip durations at the pricing structure boundaries that was identified in this paper, additional dynamic costing structures and incentives can be introduced to positively influence usage patterns. City councils can enforce modified pricing at peak hours on specified routes to alleviate pressure from congested roads by incentivizing public bicycle users through longer cost-free durations and reduced penalties for longer trip durations. A further step would be to dynamically link pricing structure to the current traffic conditions so that incentives adapt to the occurrence or severity of traffic conditions. For instance, several mapping applications, such as Google Maps and Apple Maps, provide live traffic information based on crowd-sourced data from mobile phone users on the roads. Using this data as an input, city councils can specifically lower public bicycle prices on congested car routes to increase bikeshare demand and relieve congestion along those routes. The drawback of this approach is the potential lack of predictability and visibility of incentives due to their non-periodic nature. A hybrid approach may be the most suitable with a fixed schedule for reduced costs along specific routes coupled with dynamic cost reductions at times and routes where traffic arises.

A further implication of the results here is that the expansion of cost-driven public share bicycle systems can have distinctly different requirements from convenience-driven private cycling infrastructure. For instance, the city of Amsterdam is planning nearly €200 Million of investment by 2040 to alleviate pressure off its bike infrastructure particularly in central urban regions where mainly private cyclists have indicated the highest degree of inconvenience in parking their bikes [20]. Their strategy focuses on increasing the number of bike racks in central locations where population and bike densities are highest. The current study indicates that, unlike *private* bicycle users that prioritize convenience in their transport decisions [20], *public* bicycle users can be more sensitive to monetary cost in their trip patterns. In particular, increasing the density of stations within currently covered neighborhoods to alleviate traffic in central areas of cities may not effectively increase uptake. New stations that increase the reach of the current network into new areas, where the trip times to existing stations remain within the cost-free period, are more likely to increase uptake of public bicycles and encourage new segments of society to cycle [25]. Cities that aim to increase uptake among private and public cyclists should plan new infrastructure that balances the considerations of convenience, with ample bike racks in central locations for private users, and cost, with appropriately separated stations that maximize coverage within the cost-free time limit. Retrospective measures are also possible for existing infrastructure by adjusting pricing structures, particularly the cost-free period, to generate public bicycle demand in low uptake areas.

More broadly, understanding the cost sensitivity of urban mobility and its interplay with energy budgets is relevant for the design of the costing structures of other transport systems in shaping citywide traffic patterns. The planning and design of car sharing systems [21], which involve fleets of vehicle owned by private companies, also requires meticulous consideration of the costing structure and the locations at which shared cars can be borrowed or left. In terms of topology, the emerging Personal Rapid Transport (PRT) [22], which is proposed as a public and more environmentally friendly replacement for cars, bears high resemblance to public bikes in that it captures the freedom of personal choice and explicit usage cost in mobility decisions. Characterizing the extent to which cost or distance impact routing decisions in shared transportation systems, and validating it experimentally with social incentives [23], will play a significant role in the design and expansion of this transportation paradigm.

## Conclusion

This paper has analysed the impact of cost and topology on trip duration distributions in public bicycle systems. It has confirmed earlier reports [17,27] on similarity in trip duration distributions for public bike share usage across cities, and has found that registered users in particular exhibit a high cost sensitivity around the 30 and 60 minute pricing boundaries. The analysis has also shown that the spatial topology of the bikeshare system is not a strong driver of usage patterns, with the significant differences in station and route statistics for the two cities not affecting the similarity in trip durations. It is likely that more universal factors, such as energy rather than monetary cost [27], drive the overall similarities in behavior while the pricing structure tunes behaviors within the constraints of people's daily energy budgets for movement.

## Acknowledgments

The author would like to thank Brano Kusy for useful discussions related to this study. The public bicycle data for Boston and Washington D.C. is publically available through the Hubway and CapitalBikeshare web portals respectively [14,15].

## Author Contributions

Conceived and designed the experiments: RJ. Performed the experiments: RJ. Analyzed the data: RJ. Contributed reagents/materials/analysis tools: RJ. Wrote the paper: RJ.

## References

1. Fishman E, Washington S, Haworth N (2013) Bike Share: A Synthesis of the Literature. Transport Reviews. 33(2):148–165.
2. Rhee I, Shin M, Hong S, Lee K, Chong S (2011) On the levy-walk nature of human mobility. IEEE/ACM Trans. Netw. 19(3):630–643.
3. Gonzalez MC, Hidalgo CA, Barabasi AL (2008) Understanding individual human mobility patterns. Nature, 453(7196):779–782.
4. Noulas A, Scellato S, Lambiotte R, Pontil M, Mascolo C (2012) A Tale of Many Cities: Universal Patterns in Human Urban Mobility. PLoS ONE. 7(5): e37027.
5. Bell M, Ward G (2000) Comparing temporary mobility with permanent migration. Tourism Geographies, 2(1):97–107.
6. Sjaastad LA (1962) The Costs and Returns of Human Migration. Journal of Political Economy, Vol. 70(5): 80–93.
7. Borjas GJ (1994) The economics of immigration. Journal of Economic Literature, 32(4):1667–1717.
8. Massey DS, Arango J, Hugo G, Kouaouci A, Pellegrino A, et al. (1993) Theories of international migration: a review and appraisal. Population and development review, 431–466.
9. Carrington WJ, Detragiache E, Vishwanath T (1996) Migration with Endogenous Moving Costs. The American Economic Review, 86(4):909–930.
10. Zipf GK (1946) The P1 P2/D Hypothesis: On the Intercity Movement of Persons. American Sociological Review. 11(6): 677–686.
11. Mátyás L (1997) Proper econometric specification of the gravity model, The world economy. 20(3):363–368.

12. Simini F, González MC, Maritan A, Barabási AL (2012) A universal model for mobility and migration patterns. Nature. 484:96–100.
13. Kaltenbrunner A, Meza R, Grivolla J, Codina J, Banchs R (2010) Urban cycles and mobility patterns: Exploring and predicting trends in a bicycle-based public transport system. Pervasive Mob. Comput. 6(4):455–466.
14. "Boston Data Visualisation Challenge: The Hubway Trip History Data", Metropolitan Area Planning Council, , (http://hubwaydatachallenge.org/trip-history-data/, last accessed Sep. 2013).
15. "Capital Bikeshare Trip History Data, District Department of Transportation", (http://www.capitalbikeshare.com/trip-history-data, last accessed Sep. 2013).
16. Brockmann DD, Hufnagel L, Geisel T (2006) The scaling laws of human travel. Nature. 439:462–465.
17. Padgham M (2012) Human Movement Is Both Diffusive and Directed. PLoS ONE 7(5): e37754.
18. Nair R, Miller-Hooks E, Hampshire RC, Bušić A (2013) Large-Scale Vehicle Sharing Systems: Analysis of Vélib. International Journal of Sustainable Transportation. 7(1):85–106.
19. Froehlich J, Neumann J, Oliver N (2009) Sensing and predicting the pulse of the city through shared bicycling. In Proceedings of the 21st international joint conference on Artifical intelligence (IJCAI'09), Hiroaki Kitano (Ed.). Morgan Kaufmann Publishers Inc., San Francisco, CA, , USA, 1420–1426.
20. "Summary: Long-term Bicycle Plan 2012 – 2016," City of Amsterdam Report (www.amsterdam.nl/publish/pages/458806/samenvatting_eng_lowres3.pdf, last accessed July 2013).
21. Shaheen SA, Cohen AP, Chung MS (2009) North American Carsharing: 10-Year Retrospective. Transportation Research Record: Journal of the Transportation Research Board, 35–44.
22. Gilbert R, Perl A (2010) Grid-connected vehicles as the core of future land-based transport systems, Energy Policy, Volume 35, Issue 5, May 2007, Pages 3053–3060.
23. Gaker D, Zheng Y, Walker J (2011) The Power and Value of "Green" in Promoting Sustainable Travel Behaviours. In Proceedings of World Congress on Transportation Research.
24. Shaheen S, Guzman S, Zhang A (2010) Bikesharing in Europe, The Americas, and Asia: Past, Present, and Future, Transportation Research Record, No. 2143, pp. 159–167.
25. BuckD, BuehlerR, HappP, RawlsB, ChungP, et al.Are Bikeshare Users Different from Regular Cyclists? First Look at Short-Term Users, Annual Members, and Area Cyclists in the Washington, D.C., Region – (accessed online Sep 9 2013: http://amonline.trb.org/2vege9/2vege9/1#sthash.UD03lp7S.dpuf)
26. DeMaio P (2009) Bike-sharing: History, Impacts, Models of Provision, and Future. Journal of Public Transportation Vol. 12, No.4.
27. Kölbl R, Helbing D (2003) Energy laws in human travel behaviour. New J. Phys. 5 48.
28. Social Bicycles. accessed online September 9, 2013: http://socialbicycles.com
29. Shaheen S, Martin E, Cohen A, Finson R (2012) Public Bikesharing in North America: Early Operator and User Understanding, MTI-11-26, San Jose, California, 138 pp.